\documentclass[twocolumn,showpacs,amsmath,amssymb,prl,aps]{revtex4}   %%  4-pages !!
\usepackage{graphicx}% Include figure files
\usepackage{dcolumn}% Align table columns on decimal point
\usepackage{bm}% bold math
\usepackage{color}% color text

\renewcommand{\vec}[1]{\mathbf{#1}}

\newif\ifgraph

\graphtrue             %
\begin{document}
\title{
Self-propelled Janus particles in a ratchet: Numerical simulations}

\author{Pulak~Kumar~Ghosh$^{1}$, Vyacheslav~R.~Misko$^{1,2}$, Fabio~Marchesoni$^{1,3}$, and Franco Nori$^{1,4}$}
\affiliation{$^{1}$ CEMS, RIKEN, Saitama, 351-0198, Japan}
\affiliation{$^{2}$Departement Fysica, Universiteit Antwerpen,
B-2020 Antwerpen, Belgium} \affiliation{$^{3}$ Dipartimento di
Fisica, Universit\`{a} di Camerino, I-62032 Camerino, Italy}
\affiliation{$^{4}$ Physics Department, University of Michigan,
Ann Arbor, MI 48109-1040, USA}

\date{\today}

\begin{abstract}
Brownian transport of self-propelled overdamped microswimmers
(like Janus particles) in a two-dimensional periodically
compartmentalized channel is numerically investigated for
different compartment geometries, boundary collisional dynamics,
and particle rotational diffusion. The resulting time-correlated
active Brownian motion is subject to rectification in the presence
of spatial asymmetry. We prove that ratcheting of Janus particles
can be orders of magnitude stronger than for ordinary thermal
potential ratchets and thus experimentally accessible. In
particular, autonomous pumping of a large mixture of passive
particles can be induced by just adding a small fraction of Janus
particles.
\end{abstract}
 \pacs{
82.70.Dd %Colloids
87.15.hj %Transport dynamics
36.40.Wa %Charged clusters
%02.40.-k; %Geometry, differential geometry, and topology
%74.25.Wx %Vortex pinning (includes mechanisms and flux creep)
%74.25.Uv; %Vortex phases (includes vortex lattices, vortex liquids, and vortex glasses)
%74.78.Na; %Mesoscopic and nanoscale systems
}
\maketitle

Rectification of Brownian motion has been the focus of a concerted
effort, both conceptual \cite{Reimann} and technological
\cite{RMP2009}, aimed at establishing net particle transport on a
periodic substrate in the absence of external biases. To this purpose
two basic ingredients are required (Pierre Curie's conjecture): a
spatial asymmetry of the substrate and a time correlation of the
(non-equilibrium) fluctuations, random or deterministic, applied to
the diffusing particles. Each particle is assumed to interact with
the substrate via an appropriate periodic potential, also called
ratchet potential. Typically, demonstrations of the ratchet effect
had recourse to external {\it unbiased} time-dependent drives (rocked
and pulsated ratchets); rectification induced by time-correlated, or
colored, fluctuations (thermal ratchets) seems to be of no practical
use, despite its conceptual interest.

Brownian diffusion in a narrow, corrugated channel can also be
rectified according to Curie's conjecture. The constituents of a
mixture of repelling particles in a periodically-modulated
channel, are pressed against the channel walls so that their
dynamics becomes sensitive to any asymmetry of the channel
compartments (collective geometric ratchet). Subjected to an a.c.
drive oriented along the channel axis, the mixture drifts in the
easy-flow direction, where the average compartment corrugation is
the less steep \cite{Wambaugh}, although with {\it much lower}
efficiency than in ordinary ratchet potentials. Such a collective
ratchet mechanism has been experimentally observed for a.c. drives
and relatively high particle densities \cite{vortex1,vortex2},
whereas the net current apparently vanishes at low densities
\cite{Wambaugh}. A simple kinetic equation argument
\cite{Zwanzig,ChemPhysChem} suggests that rectification of
mixtures of repelling particles, or even single particles, in an
asymmetric channel can also be induced by time-correlated thermal
fluctuations, like in thermal ratchets. However, being thermal
ratchets weak in general and (low-density) collective geometric
ratchets less performing than potential ratchets, demonstration of
such an effect seems beyond reach. On the other hand,
rectification of Brownian diffusion by an internal energy source,
like the nonequilibrium fluctuations invoked to power thermal
ratchets, is very appealing: The diffusing particles would harvest
kinetic energy directly from their environment, without requiring
any externally applied field (though unbiased), and transport
would ensue as an {\it autonomous} symmetry-directed particle
flow.

To enhance rectification of time correlated-diffusion in a
modulated channel with zero drives, we propose to use a special
type of diffusive tracers, namely of active, or self-propelled,
Brownian particles. Self-propulsion is the ability of most living
organisms to move, in the absence of external drives, thanks to
an``engine'' of their own \cite{Purcell}. Self-propulsion of
micro- and nano-particles (artificial microswimmers) poses a
challenge with respect to their unusual nonequilibrium diffusion
properties as well as their applications to nanotechnology
\cite{Schweitzer}. Recently, a new type of microswimmers has been
synthesized, where self-propulsion takes advantage of the local
gradients that asymmetric particles can generate in the presence
of an external energy source (self-phoretic effect). Such
particles, called Janus particles \cite{Chen}, consist of two
distinct ``faces", only one of which is chemically or physically
active. Such two-faced objects can induce either concentration
gradients, by catalyzing some chemical reaction on their active
surface \cite{Paxton1,Gibbs}, or thermal gradients, by
inhomogeneous light absorption (self-thermophoresis)
\cite{Sano,Bechinger} or magnetic excitation (magnetically induced
self-thermophoresis \cite{ASCNano2013JM}). Moreover, experiments
demonstrated the ability of Janus microswimmers to perform guided
motions through periodic arrays \cite{Bechinger} and separate
colloidal mixtures, due to their selective interaction with the
constituents of the mixture \cite{WeSM}.

An active microswimmer gets a continuous push from the environment,
which in the overdamped regime (inertia effects are generally
neglected) corresponds to a self-propulsion velocity ${\vec v_0}$
with constant modulus $v_0$ and direction randomly varying in time
with rate $\tau^{-1}_\theta$. In a two-dimensional (2D) boundless
suspension, the position $\vec{r}(t)=(x(t), y(t))$ of the
microswimmer diffuses according to F\"urth's law
\begin{equation}
\label{furth} \langle \Delta \vec{r}(t)^2\rangle = 4
(D_0+v_0^2\tau_\theta/4)t
+(v_0^2\tau_\theta^2/2)(e^{-2t/\tau_\theta}-1),
\end{equation}
where $\Delta \vec{r}(t) = \vec{r}(t)-\vec{r}(0)$ and $D_0$ is the
translational diffusion constant of a passive particle of the same
geometry at a fixed temperature. The mechanisms responsible for
translational and rotational diffusion are not necessarily the same
\cite{Lowen,Gibbs} and therefore $D_0$, $v_0$ and $\tau_\theta$ can
be treated as independent model parameters. The diffusion law of Eq.
(\ref{furth}) is due to the combined action of two statistically
independent 2D Gaussian noise sources \cite{Marchetti}, a
delta-correlated thermal noise, $\vec{\xi}_0(t)$ and a colored
effective propulsion noise, $\vec{\xi}_c(t)$, with correlation
functions $\langle \xi_{0,i}(t)\rangle=0$, $\langle
\xi_{c,i}(t)\rangle=0$, $\langle
\xi_{0,i}(t)\xi_{0,j}(0)\rangle=2D_0\delta_{ij}\delta (t)$, and
$\langle
\xi_{c,i}(t)\xi_{c,j}(0)\rangle=2(D_c/\tau_\theta)\delta_{ij}e^{-2|t|/\tau_\theta}$,
with $i,j=x,y$ and $D_c=v_0^2\tau_\theta/4$. Correspondingly, the
microswimmer mean self-propulsion path is $l_\theta=v_0\tau_\theta$.

When confined to a constrained geometry, like a channel, with
compartment size smaller than the self-propulsion length, $l_\theta$,
the microswimmer undergoes multiple collisions with the walls and the
confining geometry comes into play (Knudsen diffusion
\cite{Brenner}). Contrary to standard thermal ratchets in asymmetric
potentials \cite{Hanggi}, where the strength of the colored noise is
kept constant, here $D_c$ grows linearly with $\tau_\theta$ (i.e.,
the variance of ${\vec \xi_c}(t)$ is set to $v_0^2$). As a
consequence, increasing $\tau_\theta$ not only makes geometric
rectification effective even in the case of a single particle, but
also enhances the power dissipated to fuel its self-propulsion. As a
result, rectification in active Brownian ratchets can be so much
stronger than in ordinary thermal ratchets that {\it direct
observation} becomes possible. For instance, our simulations show
that even a small fraction of interacting micro-swimmers suffices to
drag along a large mixture of passive particles ({\it autonomous
pumping}).

\begin{figure}
\centering
\includegraphics[width=0.44\textwidth]{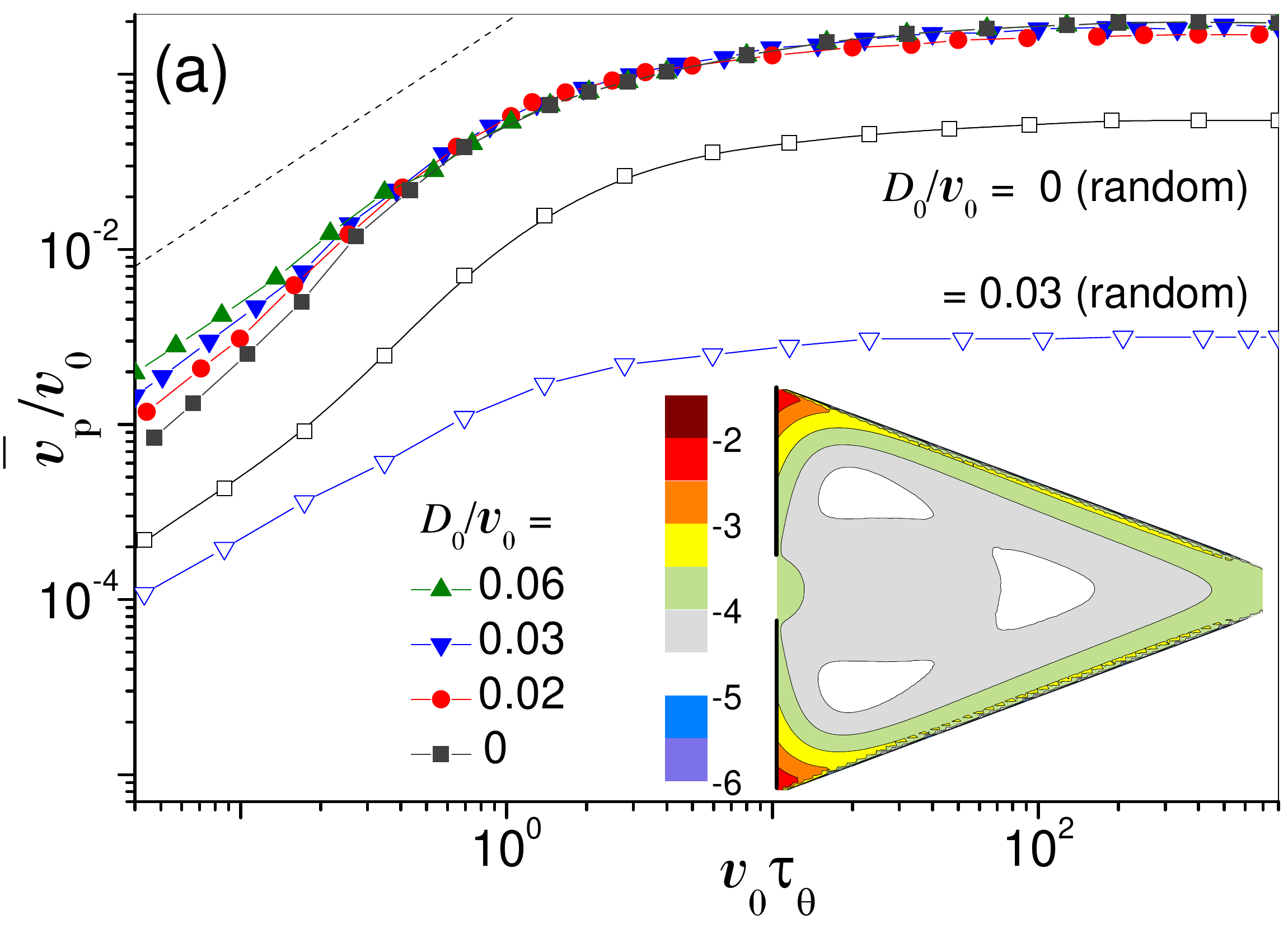}
\includegraphics[width=0.44\textwidth]{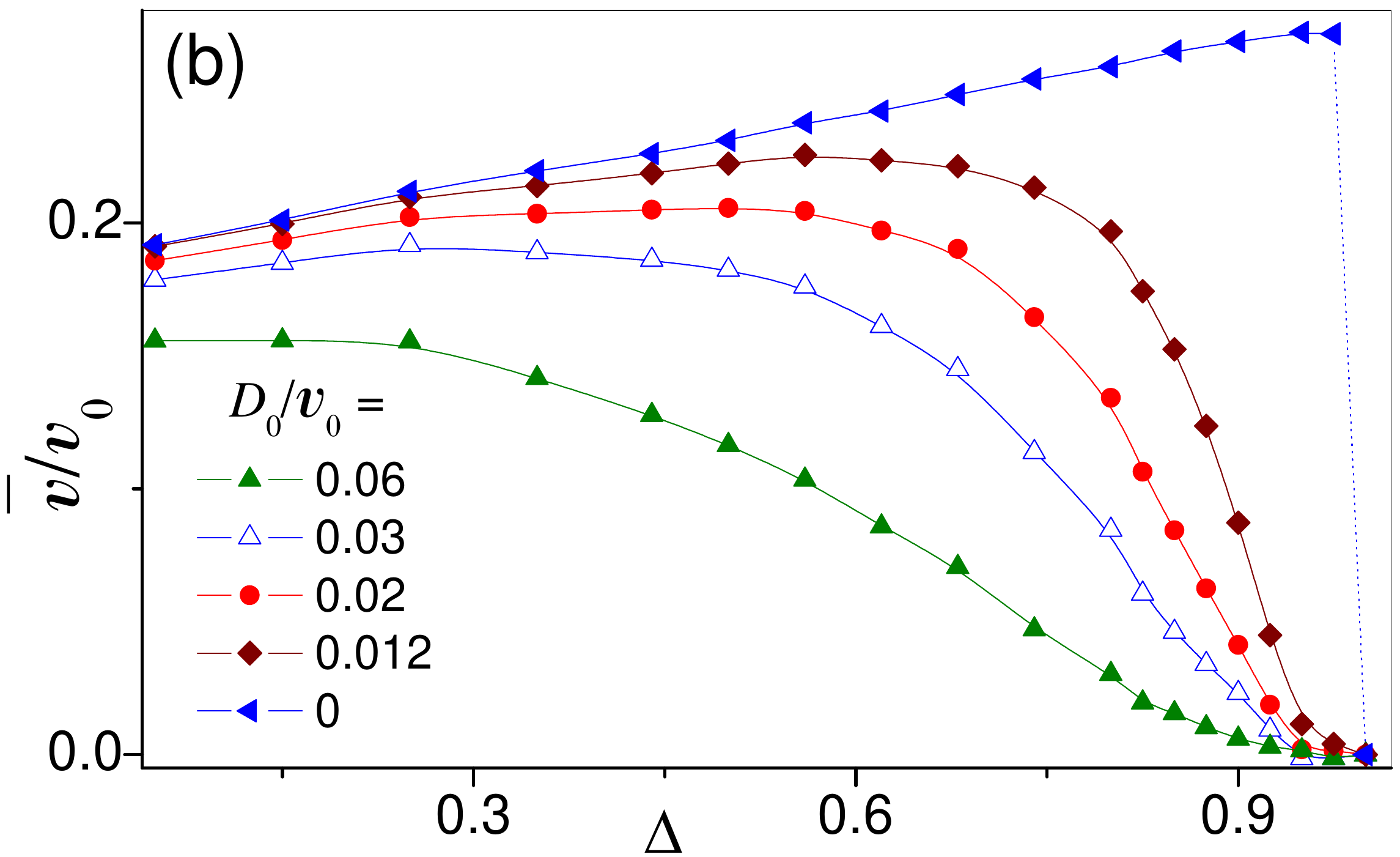}
\caption{(Color online) Rectification of a single pointlike Janus
particle with self-propulsion speed $v_0$ in a triangular channel
with compartment size $x_L=y_L=1$: (a) average velocity $\bar v$
vs. $\tau_\theta$ for channel pore size $\Delta=0.1$, different
$D_0$ and sliding (filled symbols) or randomized b.c. (empty
symbols). A dashed line with slope 1 is drawn for reader's
convenience. Inset: (logarithmic) contour plot of the stationary
probability density $P(x,y)$ in a channel compartment; (b) average
${\bar v}$ vs. $\Delta$ for different $D_0$, sliding b.c. and
$\tau_\theta=300$. \label{F1}}
\end{figure}

{\it Autonomous Janus ratchets.} --- The rectification of a Janus
particle in a 2D asymmetric channel was simulated by numerically
integrating the Langevin equations \cite{Marchetti}, $\dot
x=v_0\cos \theta +\xi_{0,x}(t)$, $\dot y=v_0\sin \theta
+\xi_{0,y}(t)$, $\dot \theta=\xi_\theta(t)$, where $\xi_{0,x}(t)$
and $\xi_{0,y}(t)$ have been defined above and $\xi_{\theta}(t)$
is an additional 1D Gaussian noise with $\langle
\xi_{\theta}(t)\rangle=0$ and $\langle
\xi_{\theta}(t)\xi_{\theta}(0)\rangle=2D_\theta\delta (t)$,
modeling the fluctuations of the self-propulsion angle $\theta$,
measured, say, with respect to the {\it positive} channel
easy-flow direction. As $\langle \cos \theta (t) \cos \theta (0)
\rangle=\langle \sin \theta (t) \sin \theta (0)
\rangle=(1/2)e^{-|t|/D_\theta}$, it follows immediately that, in
the notation of Eq. (\ref{furth}), $v_0 \cos \theta$ and $v_0 \sin
\theta$ play the role of $\xi_{c,x}(t)$ and $\xi_{c,y}(t)$,
respectively, with $\tau_\theta=2/D_\theta$. The channel
compartments were taken triangular in shape, with length $x_L$,
width $y_L$ and pore size $\Delta$ [see inset in Fig.
\ref{F1}(a)]. Throughout our simulations we kept the compartment
aspect ratio $r=x_L/y_L$ constant with $r=1$, and, by rescaling $x
\to x/\kappa$ and $y\to y/\kappa$, we conveniently set
$x_L=y_L=1$. Analogously, we rescaled $t\to v_0 t/\kappa$, so as
to work with a constant self-propulsion velocity, $v_0=1$. In
summary, the output of our integration code only depends on two
rescaled noise intensities, $D_0/\kappa v_0$ and $\kappa
D_\theta/v_0$ (or, equivalently, $v_0\tau_\theta/\kappa)$, and one
geometric parameter $\Delta/y_L$.

The collisional dynamics of a Janus particle at the boundaries was
modeled as follows. The translational velocity $\vec{\dot r}$ is
elastically reflected, whereas for the coordinate $\theta$ we
considered two possibilities: (i) frictionless collisions, $\theta$
unchanged. The microswimmer slides along the walls for an average
time of the order of $\tau_\theta$, until the noise $\xi_\theta (t)$
redirects it toward the center of the compartment. The inset in Fig.
\ref{F1}(a) clearly shows the accumulation of the stationary particle
probability density $P(x,y)$ along the boundaries; (ii) rotation
induced by a tangential friction, $\theta$ randomized. This causes
the particle to diffuse away from the boundary, which in general
weakens the rectification effect [see Fig. \ref{F1}(a)]. Note that in
the case of elastic boundary reflection of both $\vec{\dot r}$ and
$\vec{v}_0$, the microswimmer motion would amount to an ordinary
equilibrium Brownian diffusion with finite damping constant,
$\gamma=2/\tau_\theta$ \cite{inertia}, and rectification would be
suppressed.

In Fig. \ref{F1} we report our results for the rectification
current, ${\bar v} \equiv \langle \dot x \rangle$ (in units of
$v_0$), of a pointlike Janus particle in a triangular channel with
fixed compartment dimensions and varying $\tau_\theta$, panel (a),
and $\Delta$, panel (b). In panel (a) the pore size was set to
$\Delta =0.1$ and several curves ${\bar v}$ versus $\tau_\theta$
were computed for different $D_0$, i.e, at different temperatures,
and sliding boundary conditions (b.c., filled symbols). At large
$\tau_\theta$, microswimmer diffusion is of the Knudsen type and
rectification is dominated by self-propulsion; all curves ${\bar
v}(\tau_\theta)$ increase monotonously with $\tau_\theta$ until
they level off [the weak $D_0$ dependence of such asymptotes is
shown in Fig. \ref{F1}(b)]. Most importantly, we obtained ratios
${\bar v}/v_0$ in excess of 20$\%$, which means that here
\emph{the rectification power is orders of magnitude larger than
for single-particle thermal ratchets in an asymmetric potential}
\cite{RMP2009}. Moreover, the simulation parameter values adopted
here, in rescaled units, are consistent with the corresponding
values reported in the experimental literature, see, e.g., Table 1
of Ref. \cite{Bechinger}; hence, the possibility of a direct
demonstration of this striking effect.
\begin{figure}
\centering
\includegraphics[width=0.44\textwidth]{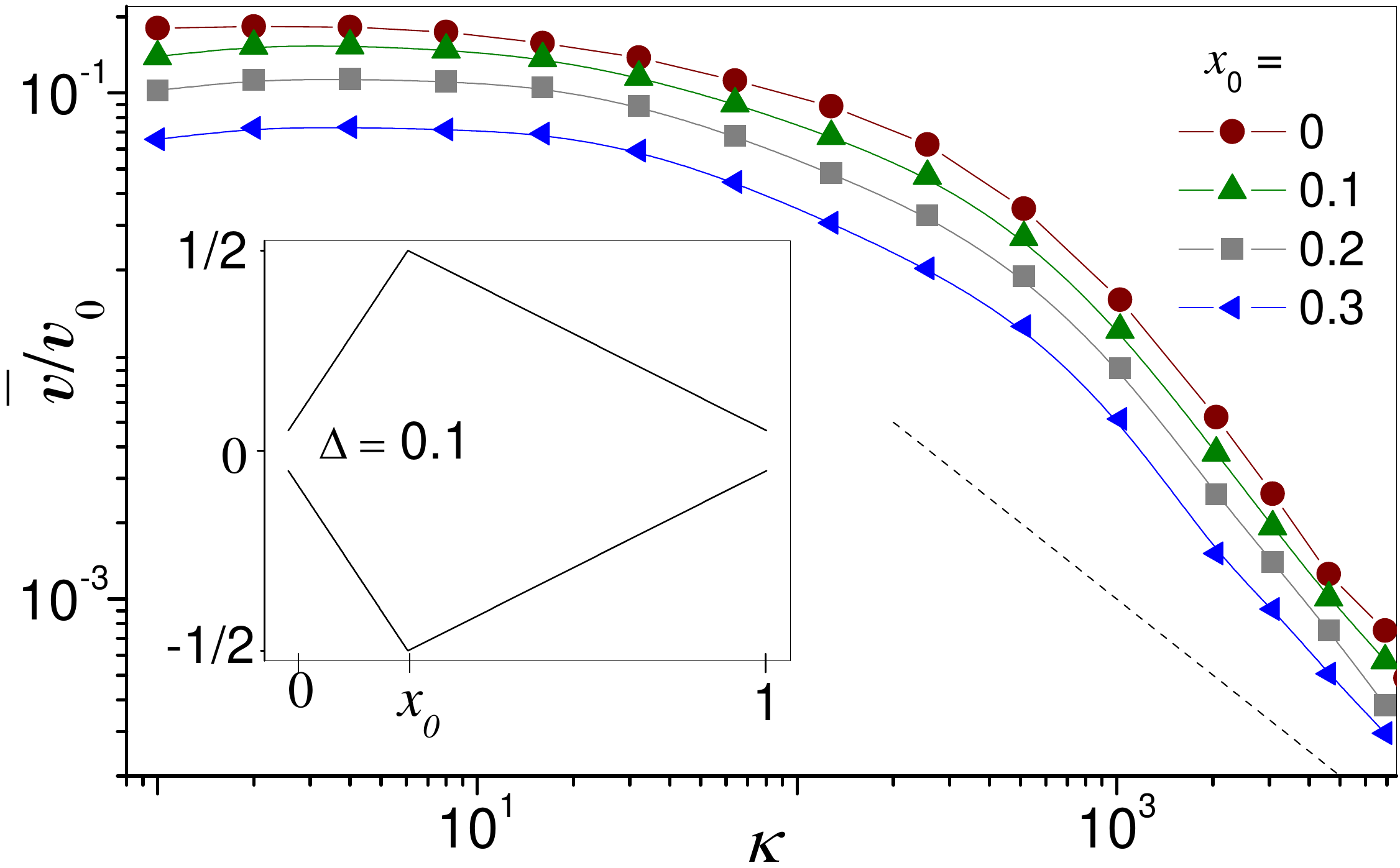}
\caption{(Color online) Rectification of single Janus particle
with $v_0=1$ in asymmetric channels with different geometries. A
typical compartment is sketched in the inset: $x_L$, $y_L$, and
$\Delta$ are as in Fig. \ref{F1}(a), i.e., $\kappa=1$, but the
corners are shifted by $x_0$. The compartment dimensions have then
be rescaled by a magnification factor $\kappa$. Main panel:
average velocity $\bar v$ versus $\kappa$ for $\tau_\theta=300$,
sliding b.c. and different $x_0$. A dashed line with slope $-1$ is
drawn for reader's convenience.\label{F2}}
\end{figure}

For the sake of a comparison in panel (a) we also report two curves
${\bar v}(\tau_\theta)$ obtained by imposing $\theta$ randomization
at the boundaries (empty symbols). As expected ${\bar v}/v_0$
decreases as the persistency of the particle self-propulsion is
suppressed by its collisions against the walls. The thermal
fluctuations ${\vec \xi}_0(t)$ further suppress rectification as they
induce more wall collisions and, thus, stronger $\theta$
randomization at the boundaries. In other words, randomized b.c. tend
to statistically couple thermal fluctuations and random
self-propulsion mechanism. Finally, we notice that the curves ${\bar
v}(\tau_\theta)$ at low $\tau_\theta$ shift upwards on raising the
thermal noise level $D_0$. This effect becomes apparent for $l_\theta
<x_L$, that is, for $\tau_\theta$ smaller than the average time a
self-propelled particle takes to exit a compartment. Under such
circumstances, thermal noise assists the rectification process.
Moreover, on taking the long-time limit of the diffusion law in Eq.
(\ref{furth}), $\langle \Delta \vec{r}(t)^2\rangle = 4 (D_0+D_c)t$,
we see that at even smaller self-propulsion lengths,
$l_\theta<4D_0/v_0$, self-propulsion can be neglected with respect to
thermal fluctuations.

Panel (b) of Fig. \ref{F1} illustrates the dependence of ${\bar
v}$ on the pore size for increasing $D_0$ at large $\tau_\theta$
and sliding b.c. Here the rectification power is suppressed by the
thermal fluctuations and is the highest for large $\Delta$. The
first trend is opposite to that observed for low $\tau_\theta$:
now ${\vec \xi}_0(t)$ helps the Janus particle bypass the
compartment corners, when ${\vec v}_0$ pushes it in the negative
direction [see inset of panel (a)]. Therefore, the diode-funneling
effect exerted by the triangular compartments can be either
enhanced or weakened by delta-correlated fluctuations, depending
on the regime of self-propulsion. The second trend might sound
counterintuitive, were not that for larger $\Delta$ and fixed
$y_L$ the sides of the channel grow shorter and the particle takes
less time to slide along them and through the exit pore. On the
other hand, the backward flow with negative velocity is blocked
mostly at the compartment corners, regardless of the actual pore
size. The moderate $\Delta$ dependence reported in Fig.
\ref{F1}(b) indicates that our numerical analysis can be safely
extended to Janus swimmers of finite radius.

Finally, in view of practical applications, we tested the robustness
of Janus particle rectification in channels with variable degrees of
asymmetry. In Fig. \ref{F2} we varied the channel geometry by
symmetrically shifting the compartment corners by a fixed amount
$x_0\in [0,0.5)$ (see inset). Moreover, we also enlarged the
compartments by a scaling factor $\kappa$, so as to accommodate for a
variety of experimental set-ups. One immediately sees that the
rectification power decreases by only a factor 2 for $x_0$ up to 0.2
and is insensitive to $\kappa$ as long as $l_\theta>\kappa x_L$. For
much larger $\kappa$, the particle spends most of its time away from
the (asymmetric) compartment walls and $\bar v$ drops inversely
proportional to $\kappa$. Indeed, the rescaled intensity of ${\vec
\xi}_0(t)$, $D_0/\kappa v_0$, is suppressed with respect to the
rescaled propulsion noise intensity, $D_c=\kappa D_\theta/v_0$, which
means that for $\kappa \to \infty$,  $\kappa {\bar v}/v_0$ tends to a
constant, that is, ${\bar v} \propto \kappa^{-1}$  [see curves for
$D_0=0$ in Fig. \ref{F1}(a).]

{\it Janus particles as autonomous pumps.}--- The remarkable
robustness of the rectification mechanism investigated here lends
itself to practical applications. Let us consider a binary mixture
consisting of $N_m$ Janus microswimmers with {\it large} $l_\theta$
(active particles) and $N_p$ non-self-propelling objects (passive
particles). For simplicity, both species are represented by soft
interacting disks of radius $r_0$ and repulsive force of modulus
$F_{i,j}=k(2r_0-r_{ij})$ if $r_{ij}<2r_0$ and $F_{i,j}=0$ otherwise
($F_{ij}$ and $r_{ij}$ denote respectively the pair force and
distance) \cite{Marchetti}. Other potentials have also been tested
with qualitatively similar results, namely, (i) active microswimmers
are capable of rectifying their motion even through a crowd of
passive particles, i.e., for packing fractions $\phi = 2\pi
r_0^2N_t/x_Ly_L$ in excess of 1. [Particles of either species overlap
in average by a length $l_o=v_0/k$ and can even pass each other];
(ii) due to their finite size and repulsive interaction with all
mixture constituents, even a low fraction of Janus microswimmers
suffices to set the entire mixture in motion ({\it autonomous
pumping}).

\begin{figure}
\centering
\includegraphics[width=0.47\textwidth]{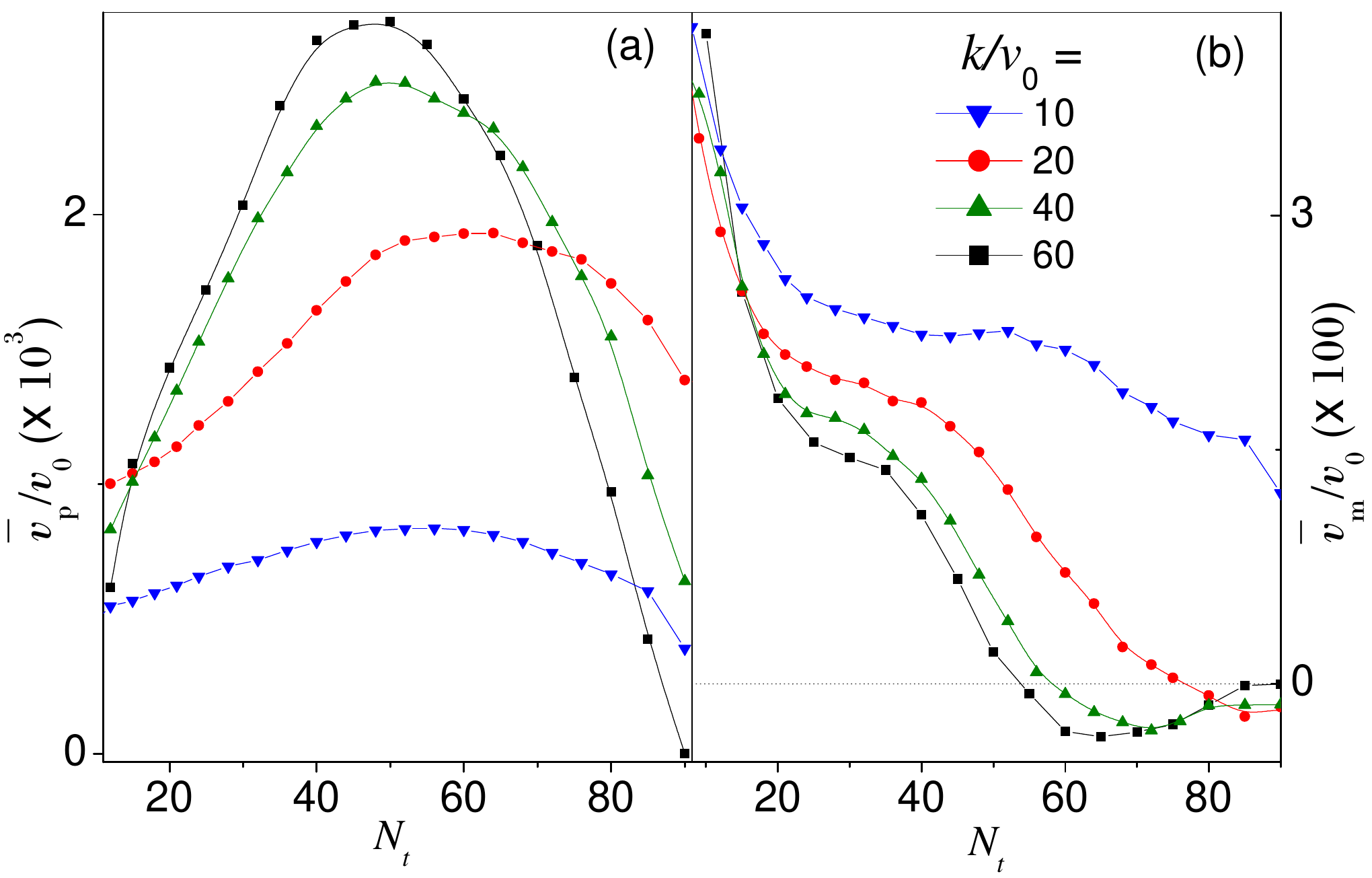}
\includegraphics[width=0.47\textwidth]{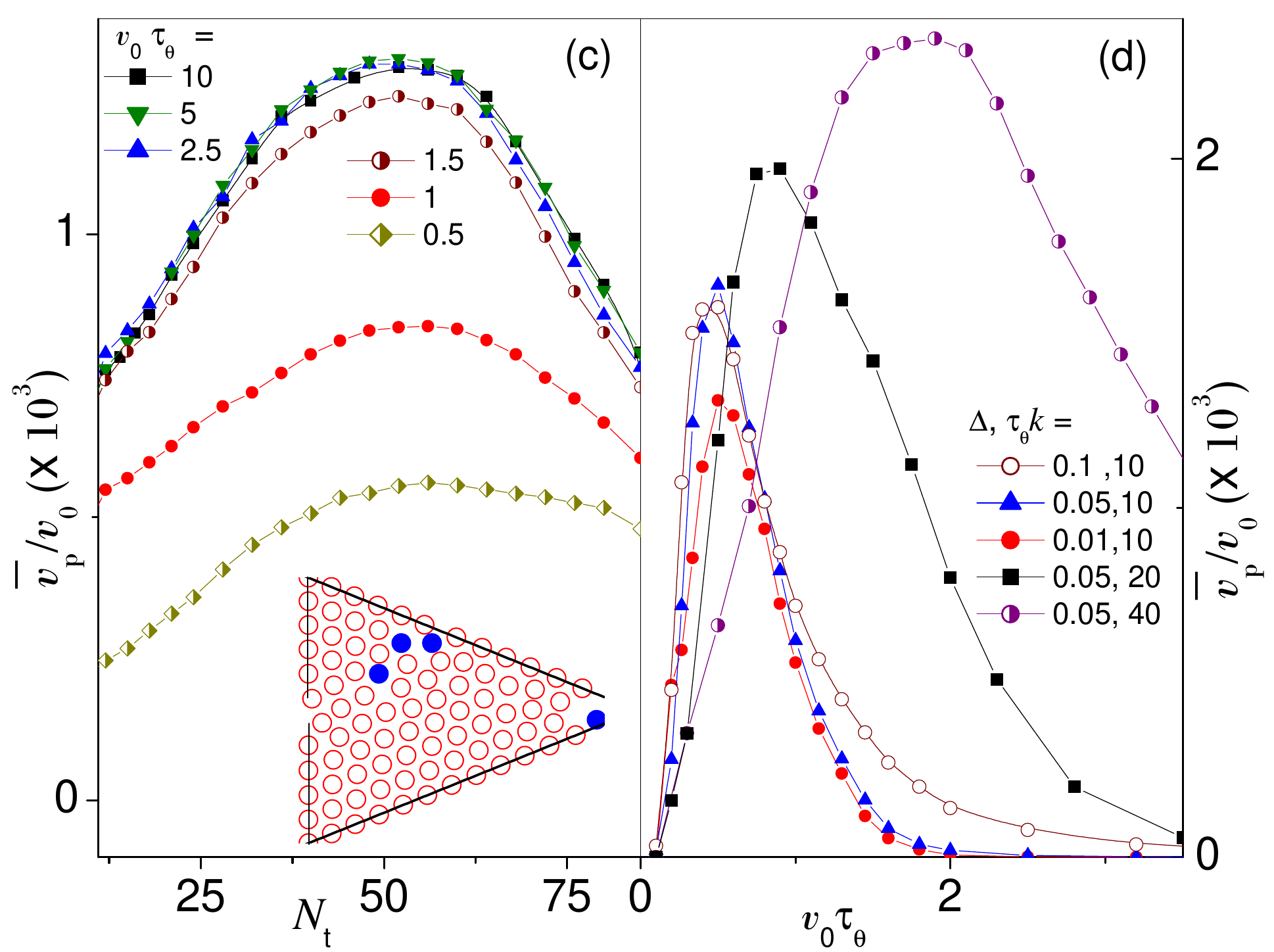}
\caption{(Color online) Rectification of a binary mixture made of
$N_m=4$ Janus particles with self-propulsion speed $v_0$, $D_0=0$,
and $N_p=N_t-N_m$ passive particles. All particles are modeled as
elastically repelling soft disks of radius $r_0=0.05$; channel
compartments are triangular with dimensions $x_L=y_L=1$. In the
inset, active and passive particles are represented by (blue) filled
and (red) empty circles, respectively. Panels (a) and (b): average
rectification velocity (a) of passive, $\bar v_p$ and (b) active
particles, $\bar v_m$, versus $N_t$ for $\Delta=0.1$ and different
interaction constant, $k$ [see legend in (b)]; (c) $\bar v_p$ vs.
$N_t$ for $\Delta=0.1$ and different $\tau_\theta$; (d) $\bar v_p$
vs. $\tau_\theta$ for $N_t=72$ and different $\Delta$ and $k$. Note
that $l_o=v_0/k$ is a measure of the disk overlap, so that
$\tau_\theta k=l/l_o$. \label{F3}}
\end{figure}

Autonomous pumping of a binary mixture of $N_t=N_m+N_p$ soft disks
was simulated under simplifying assumptions: (i) In the bulk, the
microswimmer orientation randomly changes at time intervals
$\tau_\theta$ with $\theta$ uniformly distributed in $[0,2\pi)$,
which is equivalent to setting $D_\theta=\pi^2/6\tau_\theta$; (ii)
Collisions with the walls only take place when the center of the
disks hit the boundary; disks are not elastically repelled by the
walls [see inset of Fig. \ref{F3}(c)]; (iii) Zero thermal
fluctuations, $D_0=0$, randomized b.c. (i.e., $\theta$ diffusion
is stronger close to the walls than in the bulk) were assumed for
the active microswimmer dynamics in order to demonstrate the
pumping effect under the least favorable conditions. Under sliding
b.c. pumping is appreciably stronger (not shown).

Our simulation results are reported in Fig. \ref{F3}. A few general
properties, largely independent of the model details, are discussed
here. Firstly, the mixture drifts in the easy-flow (positive)
direction, the average velocities of the active and passive particles
being, respectively, ${\bar v_m}$ and ${\bar v_p}$. In particular,
the rectification power, ${\bar v_p}/v_0$, is no smaller than
reported for potential ratchets \cite{RMP2009}. Secondly, for any
given choice of the model parameters, there exists an optimal $N_t$
where the pumping is the strongest. This defines an optimal pumping
active fraction, typically with $N_m/N_t <0.1$. Thirdly, pumping is
clearly fueled by active microswimmers as proven by $\bar v_p$
dropping for $l_\theta \lesssim x_L$ [panel (a)] and surging with
increasing the interaction constant, $k$ [panel (d)], or the fraction
of active particles (not shown). For harder disks, optimal pumping
occurs at larger $\tau_\theta$ because a Janus disk takes more time
(and a longer random path) to cross a compartment crowded with more
strongly repelling disks. As for the case of Fig. \ref{F1}(b), the
pore size $\Delta$ is clearly less important as a control parameter
than $\tau_\theta$ and $k$ [panel (d)]. Finally, we notice that for
dense mixtures, $\phi \sim 1$, the microswimmer velocity $\bar v_m$
can reverse sign. This effect was anticipated for binary mixtures on
a ratchet \cite{Savelev}: passive particles accumulate preferably on
the r.h.s. of the pore, where the cross-section is the largest;
passive particle thus act upon the diluted Janus microswimmers as an
asymmetric repulsive potential barrier, which can supersede the
funneling action of the compartment walls; hence, the Janus current
inversion at high $N_t$.

To conclude, we stress that the autonomous rectification and
pumping effects discussed here apply to biological and artificial
swimmers regardless of their propulsion mechanism, and not only to
especially fabricated Janus particles. For instance, the
autonomous robots of Ref. \cite{Viczek} are laser driven with
propulsion speed $v_0$ and about 10 $\mu$m across, which means
that their translational diffusion is negligible. The diffusion of
their propulsion orientation is due to the scattering by spatial
disorder, that is, $\tau_\theta \sim l/v_0$, where $l$ is a
disorder correlation length. In cellular systems, propulsion of
macro-biomolecules can be fueled by the ``power-stroke" associated
with the hydrolysis of ATP in suspension. In that case, $v_0\sim
\delta r/\tau_\theta$, where $\delta r$ is the net displacement
produced by a single power-stroke and $\tau_\theta^{-1}$ coincides
with the ATP hydrolysation rate in the vicinity of the biomolecule
\cite{Brenner}.

\section*{Acknowledgements} We thank RICC for computational resources.
PKG acknowledges financial support from JSPS through fellowship
No. P11502. VRM acknowledges support from the Odysseus Program of
the Flemish Government and FWO-VI. FM acknowledges partial support
from the European Commission, grant No. 256959 (NanoPower). FN was
supported in part by the ARO, JSPS-RFBR contract No. 12-02-92100,
Grant-in-Aid for Scientific Research (S), MEXT Kakenhi on Quantum
Cybernetics, and the JSPS via its FIRST program.

\end{document}

To be in a proper unit, for $\kappa =20$, corresponds to the case,

when $x_L = y_L 20 \mu m$,

$\delta = \delta - 2r_0 = 2.0 \mu m, D_0 = 0.03 \mu m^2$ second, and

$\tau_{\theta}  = 300$ second this parameter correspond to ref[].
Here our

simulation results [Fig.2] show that 10\% of self-propelled
velocity is rectified.

Again, rectification is about 20\% when $\kappa =10$.

\bibitem{Nat2009MR}
S. Metin, Nature (London) {\bf 458}, 1121 (2009).

\bibitem{APL2009Fla}
L. Zhang, J.J. Abbott, L. Dong, B.E. Kratochvil, D. Bell,
and B.J. Nelson, Appl. Phys. Lett. {\bf 94}, 064107 (2009).

\bibitem{APL2010Fla}
N. Mori, K. Kuribayashi, and S. Takeuchi,
Appl. Phys. Lett. {\bf 96}, 083701 (2010).

\bibitem{ACIE2006MM}
W.F. Paxton, S. Sundararajan, T.E. Mallouk, A. Sen,
Angew. Chem. Int. Ed. {\bf 45}, 5420 (2006).

\bibitem{ASCNano2013JM}
L. Baraban, R. Streubel, D. Makarov, L. Han, D. Karnaushenko,
O.G. Schmidt, and G. Cuniberti, ACS Nano {\bf 7}, 1360 (2013).

\bibitem{NL2008CC}
S. Sundararajan, P.E. Lammert, A.W. Zudans, V.H. Crespi, A. Sen,
Nano Lett. {\bf 8}, 1271 (2008).

\bibitem{chen}
Q.~Chen, J.K.~Whitmer, S.~Jiang, S.C.~Bae, E.~Luijten, S.~Granick,
%Q.~Chen {\it et al.},
Science {\bf 331}, 199 (2011).

\bibitem{sano}
H.-R. Jiang, N. Yoshinaga, and M. Sano,
Phys. Rev. Lett. {\bf 105}, 268302 (2010).

\bibitem{bechinger}
G. Volpe, I. Buttinoni, D. Vogt, H.-J. K\"{u}mmerer,
and C. Bechinger, Soft Matter {\bf 7}, 8810 (2011).

\bibitem{WeSM}
W. Yang, V.R. Misko, K. Nelissen, M. Kong, and F.M. Peeters,
Soft Matter {\bf 8}, 5175 (2012).

\bibitem{tune}
In addition, the swimming velocity can be controlled by
external excitation of the microswimmers which is reached
in experiments by tuning the intensity of the irradiating
laser (in case of optical excitation) or the amplitude and
frequency of the magnetic field (in case of magnetic
excitation).
Furthermore, the geometry of the channel also influences
the net flow of the passive particles, as demonstrated
in this work.

\bibitem{SMadd1}E. R. Dufresne, D. Altman, and D. G. Grier,
Europhys. Lett. \textbf{53}, 264 (2001); E. R. Dufresne, T. M.
Squires, M. P. Brenner, and D. G. Grier, Phys. Rev. Lett.
\textbf{85}, 3317 (2000).

\bibitem{SMadd2} R. D. Astumian, Phys. Rev. Lett. \textbf{91},
 118102 (2003).

\bibitem{mod}
P. H\"{a}nggi and F. Marchesoni, Rev. of Mod. Phys. \textbf{81}, 387
(2009).

\bibitem{7PRB2005Nori}
S. Savel'ev, V. Misko, F. Marchesoni, and F. Nori, Phys. Rev. B
\textbf{71}, 214303 (2005).

\bibitem{10PRB2007Kes}
K. Yu, T.W. Heitmann, C. Song, M. P. DeFeo, B.L.T. Plourde, M.B.S.
Hesselberth, and P.H. Kes, Phys. Rev. B \textbf{76}, 220507 (2007).

\bibitem{13PRL1999Nori}
J.F. Wambaugh, C. Reichhardt, C.J. Olson, F. Marchesoni, and F.
Nori, Phys. Rev. Lett. \textbf{83}, 5106 (1999).

\bibitem{4Scinece2003}
J.E. Villegas, S. Savel'ev, F. Nori, E.M. Gonzalez, J.V. Anguita, R.
Garc\'{i}a, and J.L. Vicent,
Science \textbf{302}, 1188 (2003).

\bibitem{highTC2004}
R.~W\"{o}rdenweber, P.~Dymashevski, and V.R.~Misko,
Phys. Rev. B \textbf{69}, 184504 (2004).

\bibitem{8PRB2005Vincent}
J.E. Villegas, E.M. Gonzalez, M.P. Gonzalez, J.V. Anguita,
and J.L. Vicent,
Phys. Rev. B \textbf{71}, 024519 (2005).

\bibitem{PRL2005Tonomura}
Y. Togawa, K. Harada, T. Akashi, H. Kasai, T. Matsuda, F. Nori,
A. Maeda, and A. Tonomura,
Phys. Rev. Lett. \textbf{95}, 087002 (2005).

\bibitem{9PRL2005Moshchalkov}
J. Van de Vondel, C.C. de Souza Silva, B.Y. Zhu, M. Morelle, and
V.V. Moshchalkov,
Phys. Rev. Lett. \textbf{94}, 057003 (2005).

\bibitem{3Nature2006}
C.C. de Souza Silva, J. Van de Vondel, M. Morelle, and V.V.
Moshchalkov,
Nature \textbf{440}, 651-654 (2006).

\bibitem{12PRB2010Mashchalkov}
B.B. Jin, B.Y. Zhu, R. W\"{o}rdenweber, C.C. de Souza Silva, P.H.
Wu, and V.V. Moshchalkov,
Phys. Rev. B \textbf{81}, 174505 (2010).

\bibitem{11PRB2010Plourde}
K. Yu, M.B.S. Hesselberth, P.H. Kes, and B.L.T. Plourde, Phys. Rev.
B \textbf{81}, 184503 (2010).

\bibitem{PRB2011Misko}
N.S.~Lin, T.W.~Heitmann, K.~Yu, B.L.T.~Plourde, and V.R.~Misko,
Phys. Rev. B \textbf{84}, 144511 (2011).

\bibitem{15ref1Nori}P. H\"{a}nggi, F. Marchesoni, and F. Nori, Ann.
Phys. (Leipz.) \textbf{14}, 51-70 (2005).

\bibitem{15ref2Marchesoni}P. H\"{a}nggi and F. Marchesoni, Chaos
\textbf{15}, 026101 (2005).

\bibitem{15NMat2006Nori}
D. Cole, S. Bending, S. Savel'ev, A. Grigorenko, T. Tamegai, and F.
Nori, Nature Materials \textbf{5}, 305-311 (2006).

\bibitem{Kline}
T. R. Kline, W. F. Paxton, T. E. Mallouk, and A. Sen,
Angew. Chem. Int. Ed. {\bf 44}, 744 (2005).

\bibitem{Tierno}
P. Tierno, R. Albalat, and F. Sagu\'{e}s,
Small {\bf 6}, 1749 (2010).

\bibitem{DoyleLgM22}
R. Haghgooie, C. Li, and P. S. Doyle,
Langmuir {\bf 22}, 3601 (2006).

\bibitem{LeidererPRL97}
M. K\"{o}ppl, P. Henseler, A. Erbe, P. Nielab, and P. Leiderer,
Phys. Rev. Lett. {\bf 97}, 208302 (2006).

\bibitem{KwintenEPL80}
K. Nelissen, V. R. Misko, and F. M. Peeters,
EuroPhys. Lett. {\bf 80}, 56004 (2007).

\bibitem{MiskoPRE80}
D. V. Tkachenko, V. R. Misko, and F. M. Peeters,
Phys. Rev. E {\bf 80}, 051401 (2009).

\bibitem{BradyPRL}
U. M. C\'{o}rdova-Figueroa and J. F. Brady,
Phys. Rev. Lett. {\bf 100}, 158303 (2008).

\bibitem{BradyPRL2009}
U. M. C\'{o}rdova-Figueroa and J. F. Brady,
Phys. Rev. Lett. {\bf 103}, 079802 (2009).

\bibitem{Lowen1}
B. ten Hagen, S. van Teeffelen, and H L\"{o}wen,
J. Phys.: Condens. Matter {\bf 23}, 194119 (2011).

\bibitem{Lowen2}
S. van Teeffelen and H L\"{o}wen,
Phys. Rev. E {\bf 78}, 020101(R) (2008).

\end{references}

\end{document}